\documentstyle[aaspptwo]{article}                  
\include{macros}
\def\cm{~cm$^{-1}$}
\def\mic{$\mu$m}
\def\etal{et al.\/ }
\def\odd{^{\circ}}
\begin{document}
\title{A new multiplet table for Fe I}
\author{ G.~Nave\altaffilmark{1}$^,$\altaffilmark{2}
                 \altaffiltext{1}{Blackett Laboratory, Imperial College,
                 Prince Consort Rd, London, U.K SW7 2BZ.}
                 \altaffiltext{2}{ESA external fellow}
and S.~Johansson\altaffilmark{3}\altaffiltext{3}{Lund Observatory, Box 43,
                                                  S-221 00 Lund, Sweden}
       }
\affil{Department of Physics, University of Lund, S\"olvegatan 14,
       S-223 62 Lund, Sweden}
\author{R.~C.~M.~Learner and A.~P.~Thorne}
\affil{Blackett Laboratory, Imperial College,Prince Consort Rd,
            London, U.K SW7 2BZ}
\and
\author{J.~W.~Brault\altaffilmark{4}\altaffiltext{4}{The NSO is
                          operated by the Association of Universities
                          for Research in Astronomy, Inc., under a
                          cooperative agreement with the National
                          Science Foundation. }}
\affil{National Solar Observatory, Tucson, Arizona, U.S.A }

\begin{abstract}
We have recorded spectra of iron-neon and iron-argon hollow cathode
lamps in the region 1700~\AA\ -- 5~\mic\ (59\,000 -- 2000~\cm), with
Fourier transform (FT) spectrometers at the National Solar Observatory,
Tucson, Arizona, U.S.A. and Imperial College, London, U.K., and with a
high resolution grating spectrograph at the National Institute of
Standards and Technology, Gaithersburg, U.S.A. The uncertainty of the
strongest lines in the FT spectra is $<$0.002~\cm\ (0.2~m\AA\ at 3000~\AA;
8~m\AA\ at 2~\mic\ ). Pressure and current-dependent shifts are
$<$0.001~\cm\ for transitions between low lying levels, increasing to
0.006~\cm\ for transitions between the most highly excited levels.
We report 28 new energy levels of Fe I and revised values of another 818
levels. We have identified 9501 lines as due to 9759 transitions in Fe
I, and these are presented in the form of a new multiplet table and
finding list. This compares with the $\sim$5500 lines due to 467 energy
levels in the multiplet tables of Moore (1950, 1959). The biggest
increase is in the near ultraviolet and near infra-red, and many of the
new lines are present in the solar spectrum. Experimental log(gf) values
are included where they are available. A further 125 unidentified lines
due to Fe I are given. The tables are also available in
computer-readable form.
\end{abstract}

\section{Introduction}

Spectra of low charge states of iron play a significant role in solar
and stellar spectroscopy due to the high cosmic abundance of iron. The
atomic structure of the iron-group elements is complex, and departures
from any pure coupling scheme result in a high density of lines in all
wavelength regions. The number of identified iron lines in the solar
spectrum exceeds that of any other element, and the prediction that many
unidentified solar lines in the near ultraviolet and near infrared are
almost certainly due to Fe I (Johansson 1987) has been recently verified
(Nave \& Johansson 1993b). An updated analysis of the spectrum of neutral iron,
Fe~I, has been a long-standing request from astronomers, along with the
needs for more and better laboratory data for many other elements, as
emphasised at several recent meetings (for example Smith \& Wiese 1992,
Grevesse \& Noels 1993, and the contributions of Kurucz and Lambert in
Leckrone \& Sugar 1993 ). The development of laboratory light sources
and the techniques of Fourier transform (FT) spectrometry and high
resolution grating spectrometry have made it possible to record the
spectrum with an accuracy and completeness not previously achievable.
This has enabled us to extend the term system considerably
and~to~identify~many~new~lines~in~astrophysical~spectra \\ (Nave \& Johansson,
1993a,1993b). It is now possible to present an analysis of the spectrum
which fulfills many current demands of astronomy.

Despite their age, the Revised Multiplet Table of Moore (1959) and the
Ultraviolet Multiplet Table (Moore 1950) (both of which we shall refer
to as MT) remain among the most widely-used sources of data for users
requiring extensive line lists of atomic spectra (Edl\'en \& Martin
1988). This popularity is undoubtedly due to the accuracy with which
Moore prepared them, the care with which she selected data to be
included, and the convenient format in which the data are presented.
However, modern astronomical observations supersede old ones in
signal-to-noise ratio, resolution, wavelength accuracy and spectral
wavelength coverage. The line-by-line analysis of solar and stellar
spectra has been replaced by comparisons between synthetic and observed
spectra. All this puts new demands and requirements on the atomic data
needs, and extensions and revisions to the MT have become an urgent
necessity. Extensions are particularly needed in the near ultraviolet,
where many new lines have been identified since the publication of the
UV tables, and the near infrared above 1.3~\mic, which is not covered by
the MT. Revisions of the MT are required in all wavelength regions, as
the accuracy of many lines in the MT has been substantially improved in
recent years. The requirements of astronomy now demand wavelengths to
within 1~m\AA\ in the visible and ultraviolet, and 0.001\cm\ in the
infrared. The data in the MT do not fulfill these requirements -- as a
recent search for the antecedents of a weak Fe I line showed, some of
the wavelengths quoted actually go back to measurements made over a
hundred years ago (Learner, Davies \& Thorne 1991).

In addition to the MT, Moore published three volumes of atomic energy
levels (Moore 1971), which also became standard references for atomic
spectra. These were updated for the iron-group elements by Sugar and
Corliss in 1985 (hereafter AEL-SC). The table for Fe I in the AEL-SC
incorporates the revised levels of Crosswhite (1975) and Litz\'en (1976)
and contains a list of references to earlier work. Since 1985, revisions
to the energy levels and extensions to the term system have been
made by Johansson (1987), Brown \etal (1988), O'Brian \etal (1991),
Zhu and Knight (1992); further extensions based on the spectra presented
in the current paper have been published by Johansson and Learner
(1990), Nave and Johansson (1993a,b), Nave \etal (1994), and Johansson
\etal (1994). Bi\`emont \etal (1985) have also compiled a list of Fe I
lines in the infrared solar spectrum.

The most recent compilation of gf-values for the iron-group elements is
by Fuhr, Martin \& Wiese (1988). The table for Fe I includes gf-values
for $\sim$1950 lines, some of which have an uncertainty of only a few
per cent. More recent measurements include those of O'Brian \etal
(1991), Bard, Kock \& Kock (1991), Kock, Kroll \& Schnehage (1984), and
Meylan \etal (1993). In total, roughly 2600 Fe I lines now have measured
gf values. The majority of these are between 3000~\AA\ and 1~\mic, where
almost half of the observed Fe I lines now have measured gf-values. In
the region below 3000~\AA\ this drops to only 20\%, and there are very
few lines above 1~\mic\ with measured gf-values.

Many astrophysicists now use the calculations of semi-empirical gf
values by Kurucz (1989), which are based on experimental energy levels
such as the ones presented in the present study. The calculations have
the advantage that they include significantly more lines than it is possible to
measure in the laboratory, many of which are important in analyses of
Fraunhofer spectra of the sun and cool stars. The disadvantage of these
calculations is that their accuracy cannot be estimated and is often
poor, especially for weak lines that may be the lines of most
interest in studies of the iron abundance in the sun.

In this paper we present the analysis of the laboratory spectrum of Fe
I in the form of a new multiplet table. The analysis is based on both FT
and grating measurements of iron-neon and iron-argon hollow cathode
lamps in the region 1700~\AA\ to 5~\mic\ (58\,800 $-$ 2000 \cm\ ). The
precision of the wavenumbers obtained from an FT spectrometer is up to
an order of magnitude better than that obtainable with a grating
spectrometer, but we have also included Fe I lines which were present in
only the ultraviolet grating spectra, in which the number of Fe I lines
is significantly higher (Nave \& Johansson 1993a). We present revised
values for 818 energy levels and values of 28 new levels. These give
9759 identifications of 9501 lines, which have been ordered into 2785
multiplets. Experimental gf-values are also given where they are
available. A further 125 unidentified lines due to Fe I are presented.

\section{Experimental Details}

The laboratory spectra used in this study were obtained using three
different instruments: the f/55 IR-visible-UV FT spectrometer at the
National Solar Observatory (NSO), Tucson, Arizona for the region
2000~\cm\ to 35\,000~\cm\ (5--0.29~\mic); the f/25 vacuum  UV FT
spectrometer at Imperial College, London, for FT spectra in the region
33\,000~\cm\ to 59\,000~\cm\ (3000--1700~\AA); and the 10.7m grating
spectrograph at the National Institute of Standards and Technology
(NIST), Gaithersburg, for high-dispersion grating spectra above
30\,770~\cm\ ($<$3250~\AA ).

The light source used for the FT investigations was a hollow cathode
lamp, run in either neon or argon. In addition to Fe~I spectra, this
source also gives spectra of Fe~II and the neutral and singly-ionised
spectra of the carrier gas used. The cathode was usually an uncooled
open-ended cylinder of pure iron, 8~mm in bore and 35~mm long. The metal
case of the lamp formed the anode. The running pressures were about
5~mbar of Ne or 4~mbar of Ar for the visible and IR observations, and
3--4~mbar of Ne  for the UV observations. The currents ranged from
320~mA to 1.1~A. In general, Ne gave a better signal to noise ratio for
Fe I except in the region 14~000 -- 17~500~\cm\ (7100 -- 5700~\AA ),
where the strong Ne lines raise the noise level in the spectrum.
Argon-iron spectra were therefore used in this region and were also
recorded in the region 17~500 -- 35~000~\cm\ (2857 -- 5714~\AA) to give
an absolute wavelength calibration  based on Ar II lines. One infrared
spectrum was recorded with a water-cooled cathode as source and a higher
current of 1.4~A. Survey spectra were taken out to 600~\cm\ (16\mic),
but they showed hardly any Fe I lines because of the high noise level
from the thermal infrared background.

The wavenumber, integrated intensity, width and damping of all lines in
the FT spectra were obtained using Brault's DECOMP program (Brault \&
Abrams, 1989), which fits a Voigt profile to each line. The wavenumber
scale was calibrated from 26 Ar II lines in the visible (Learner \&
Thorne 1988), and the calibration was carried into the UV and infra-red
using wide-range spectra (Nave \etal 1991, 1992). Deuterium and tungsten
lamps were used for intensity calibration.

The Doppler widths (half width at half maximum) varied from around
0.012~\cm\ at 5000~\cm\ (0.05~\AA\ at 2~\mic) to around 0.12~\cm\ at
50~000~\cm\ (5~m\AA\ at 2000~\AA). This corresponds to a Doppler
temperature of about 2500~K; the source is however, not in thermal
equilibrium, and the intensity distribution is generally quite different
from that of astrophysical spectra. Lines emitted from levels of high
excitation are also Lorentz broadened, so the observed line widths are
often higher than this. The uncertainty of the wavenumber of each line
is the sum of statistical and systematic errors. The statistical error
is equal to the half width at half maximum divided by the
signal-to-noise ratio (Brault 1987). For lines with a  signal-to-noise of
$\sim$100 this
varies from around 0.0002~\cm\ (0.8~m\AA\ at 2~\mic) in the infra-red to
0.001~\cm\ (0.01~m\AA\ at 3000~\AA) in the ultraviolet. The
weakest lines in the spectra have a signal-to-noise of $\sim$3, and the
accuracy is then of the order 0.005~\cm\ (0.02~\AA\ at 2~\mic) in the
infrared and 0.05~\cm\ (5~m\AA\ at 3000~\AA) in the ultraviolet. The most
important systematic errors are the calibration error for each spectrum
which is of the order 0.001~\cm, and possible pressure or
current-dependent shifts. The latter have been estimated at $<$0.001~\cm\ for
levels of low excitation, rising to $\sim$0.006~\cm\ for the
highest excitation levels. A full description of the procedure, with
details of the FT spectra used in the present investigation, has been
given elsewhere (Learner \& Thorne, 1988; Nave \etal, 1991; Nave \etal,
1992).

Grating spectra have been recorded in the region 30\,770 -- 58\,820~\cm\
(3250~--~1700~\AA) using iron-neon and iron-argon hollow cathode lamps.
The iron-neon hollow cathode was run in continuous mode with DC
currents of 0.6--0.8~A and a gas pressure of $ \sim$1.3~mbar. The
iron-argon hollow cathode was run in pulsed mode with peak currents of
$\sim$100~A, pulse width of 70~$\mu$\/s and frequency 100~Hz, and pressure
of $\sim$0.3--0.4~mbar. The spectra were calibrated from Ritz
wavelengths obtained from interferometrically-measured Fe II lines
(Norl\'en, private communication), and the uncertainty in the grating
wavelengths is approximately 3~m\AA\ ($\sim$~50~mK at 2500~\AA). The grating
spectra are being used in an extended and comprehensive analysis of Fe
II (Johansson \& Baschek 1988).

\section{Identification of Lines from Known Levels}

The procedure we have followed to identify Fe I lines is to compare
wavenumbers of observed lines with wavenumbers derived from measured
energy levels. We will call the latter `Ritz wavenumbers'. The energy
levels we have used in identifying the lines were from AEL-SC (Sugar \&
Corliss 1985),  Brown \etal (1988), O'Brian \etal(1991), together with
those we have located in our recent investigations of Fe I (Johansson \&
Learner
1991, Nave \& Johansson 1993a, Nave \etal 1994, Johansson \etal 1994).
To minimise the spurious coincidences of observed wavenumbers and Ritz
wavenumbers, it is important to have a set of energy level values that
is both accurate and appropriate to the sources used (O'Brian \etal
1991). We have therefore revised the energy levels from the observed
transitions in the FT spectra, using a least squares fitting program
(Radziemski \etal 1972), in which each line was assigned a weight
inversely proportional to the square of the wavenumber uncertainty.

The values of the revised energy levels are listed in table 1. We have
listed the terms in order of the energy of the lowest fine structure
level of each term. With the exception of a few levels, which we have
carefully checked, the values are within the estimated errors of those
previously published. In addition, 28 new energy levels have been
found that have not been published elsewhere, and these are marked with
an asterisk beside the level value in column 4.

Many energy levels in Fe I cannot be adequately described in any
particular coupling scheme. We have given LS designations to all the
levels where no other coupling scheme is obvious in order to label them
and to help identify which configurations are incompletely represented.
JK designations have been assigned to levels due
to the configurations $3d^64s(^6D)\,6d, 4f, 5g$ and $3d^7(^4F)4f$.
Abbreviated term designations are given in column 2. These are
extensions of the term designations used in the MT and follow the
convention established in earlier papers (Brown \etal 1988; Nave \& Johansson
1993a). The term designations and abbreviations are discussed in detail
in section 4.1. Nine of the levels could not be assigned to a
configuration and are represented only by their energy level values.

The primary criterion for identification of the lines was coincidence
between the observed wavenumber and the Ritz wavenumber derived from the
energy levels in Table 1, taking into account the uncertainty of the
wavenumbers, the uncertainty of the levels, and possible blended lines.
In many cases more than one identification is possible, particularly in
the ultraviolet where many of the lines were detectable only in the less
accurate grating spectra. Further criteria were thus applied to verify
the identifications. In the grating UV spectra, Fe I and Fe II lines
could often be distinguished by their relative intensities in spectra
obtained from pulsed and unpulsed hollow cathodes. The absorption data
of Brown \etal (1988) were also helpful as they contain no Fe II lines.
Comparison was also made with published linelists and atlases of the
solar spectrum (Moore, Minnaert \& Houtgast 1966; Moore, Tousey \& Brown
1982, 1992; Pierce and Breckinridge 1973; Swensson \etal 1970;
Livingstone and Wallace 1991; Geller 1992),
where all Fe I lines are present but Fe II lines of high excitation are
absent.

\section{Term Structure and Transitions}

The term structure of the iron-group elements is described in detail in
Johansson and Cowley (1988) and specific details pertaining to Fe I are
given in Brown \etal (1988), Johansson and Learner (1991), and Nave and
Johansson (1993a). The majority of the Fe I levels belong to the two
configuration systems $3d^64snl$ or $3d^7nl$. Terms at low excitation
energies belong to the subconfigurations $3d^6(^ML)4s4l$ and
$3d^7(^ML)4l$, and are often fairly well described in the LS coupling
scheme. Higher excitation terms are usually of the form $3d^64s(^6D)nl$
or $3d^7(^4F)nl$ and are in intermediate or JK coupling. The strongest
transitions occur within each system when the parent or grandparent
term, $^ML$, in the subconfiguration is unchanged.

\subsection{Notation of Terms}

In the MT the spectroscopic terms were abbreviated, as it was not
possible at that time to describe them by their full spectroscopic
notations. Since that time the convention used for the abbreviations has
become well established and is the one adopted in the AEL-SC. The
convention labels the lowest even-parity LS term of multiplicity M and
orbital angular momentum L as
$a\,^M\!L$, the next lowest, $b\,^M\!L$ and so on (e.g. the lowest $^3F$
term is labelled $a\,^3\!F$, the next lowest $b\,^3\!F$ \ldots). The
lowest odd-parity term is labelled $z\,^M\!L\odd$, the next lowest
$y\,^M\!L\odd$ and so on. We have kept to this convention for all the
terms that are given in the MT, but it cannot in general be extended to
higher terms. It is not consistent -- for example the lowest $^5\!D$
term is labelled $a\,^5\!D$, but the next lowest is labelled $e\,^5\!D$
-- and it is a useful designation only in LS coupling, which does not
apply to highly-excited terms.

All of the known highly-excited terms in Fe I with $n>4$ are formed by
adding a running electron to either the $3d^64s\ ^6D$ term or the $3d^7\
^4F$ term in Fe II. These two terms are called the `parent terms'.
Addition of one electron to each of these parents give the subconfigurations
$3d^64s(^6D)nl$ and $3d^7(^4F)nl$, and the abbreviations we have
adopted are intended to indicate the parent of the subconfiguration.
Terms of the form $3d^64s(^6D)nl\ ^M\!L$ are abbreviated to the form $s\
^6\!Dnl\ ^M\!L$ and terms of the form $3d^7(^4F)nl\ ^M\!L$ to $^4Fnl\
^M\!L$. Highly-excited terms with $n=4$ which were not labelled by Moore
belong to other parents and grandparents, and are of the form
$3d^6(^{M'}\!L')4s4p(^3P)\ ^M\!L$, $3d^6(^{M'}\!L')4s4p(^1P)\ ^M\!L$,
or $3d^7(^{M'}\!L')4p\ ^M\!L$. These have been given the abbreviations
$L'sp3\,^ML$, $L'sp1\,^ML$ and $^{M'}\!L'4p\,^ML$, respectively.

Many highly-excited configurations with $l\ge3$ are best described in
the JK coupling scheme rather than the LS coupling scheme. In Fe I
this applies to the subconfigurations $3d^64s(^6D)\ 6d, 4f$, $5g$ and
$3d^7(^4F)4f$. In JK coupling the levels split into groups, and each
group has as a parent the same fine structure level in Fe II (Johansson
\& Learner 1990). The J-value of this parent level ($J_c$) is coupled to
the angular momentum $l$ of the running electron to give a resultant
$K$, which is in turn coupled to the spin of the running electron to
give the J-value of the level. In Fe I, $J_c$ and $K$ are half-integral
and  $J$ is, as usual, integral. The full notation is, for example,
$3d^64s(^6D_{J_c})4f\ [K]_J$. The notation we use for the abbreviated
term designation is $s^6D_{J_c}4f\ [K]$, with similar designations for
other JK coupled subconfigurations. In the computer-readable file, $J_c$
and K have been truncated to integers. For example, the two levels due
to $3d^64s(^6D_{9/2})4f[\frac{11}{2}]$ are referred to as
$s\,^6D_{4.5}4f[5.5]$ in the printed version of the table, and
s6D4~4f[5] in the computer-readable version.

\subsection{Transitions and Selection Rules}

As a result of the mixed coupling in Fe I
the only selection rules that can reliably be applied are $\Delta J=
0,\pm 1$ ($J\!=\!0\not\rightarrow J\!=\!0$) and change in parity (we do
not see any parity forbidden lines in our laboratory spectra). However,
the multiplets containing the strongest lines fall in the visible and
follow traditional LS rules -- the transitions are between levels of the
same multiplicity, with $\Delta L = 0, \pm 1$ (but not $L=0\rightarrow
L=0$). Amongst these, the strongest lines are the ones involving the
levels with the highest J where the J-value changes in the same
direction as the L-value --- for example in multiplet $a\,^5\!F -
z\,^5\!D$ (number 88 in table 2, or MT multiplet 15) the strongest
lines correspond to J changing by -1 (sometimes called the `main diagonal'),
with the strongest line being J=5 to J=4. The next strongest are for
$\Delta J=0$ (sometimes called the `first satellites'), and the weakest
for $\Delta J=+1$ (sometimes called the `second satellites'). These
rules are not always apparent from our intensities in Table 2,
especially for multiplets involving the lowest terms ($a\,^5\!D$ and
$a\,^5\!F$) which may be affected by self absorption.

However, many multiplets do not follow LS rules, even if the levels are
of relatively low excitation. Transitions are seen both between levels
with different multiplicities and with $\Delta L>1$. The former are
sometimes called `intercombination lines', `spin-forbidden lines', or
`intersystem lines', and their presence indicates the breakdown of
the LS model. In complex spectra like Fe I they may be as common and
strong as spin-allowed transitions. Multiplets of
intercombination lines may not contain all of the theoretically
predicted transitions between the two terms, and the intensities are
often irregular as they depend on the degree of mixing with levels of
other multiplicities. For example, in multiplet $ a\,^3\!H - z\,^5\!G$
(464, MT 168) the strongest lines are $6-5$, $5-4$ and $5-5$. The $5-6$
line is not observed and the $6-6$ line is relatively weak. This
multiplet is caused by mixing between the $z\,^5\!G\odd$ term and the
$z\,^3\!G\odd$ term, and the intensities are similar to those observed
in multiplet $a\,^3\!H-z\,^3\!G\odd$ (465, MT 169). The lines to
$z\,^5\!G_6\odd$ are weak because there is no level with $J=6$ in the
$z\,^3\!G\odd$ term with which it can mix. Some multiplets are
seen in which the multiplicity changes by 4. An example is the multiplet
$a\,^1\!G - w\,^5\!G\odd$ (963, MT 517), in which both the upper
term and the lower term are mixed. In many cases, and in particular for
highly-excited levels, it is difficult to identify exactly the way in
which the levels are mixed, and a particular level may be described as a
combination of several different LS components which all contribute
more-or-less equally.

In highly-excited levels, where the dominant LS component may contribute
only 30\%, the concepts of an LS term and a multiplet have in
general no meaning. For the sake of consistency, we have arranged all the
transitions into multiplets, assigning LS labels according to the
largest LS component, but the majority of lines in the infra-red are
better arranged into `allowed supermultiplets' which consist of all the
transitions between two subconfigurations built on the same parent term.
These supermultiplets usually obey the rule $\Delta l = \pm 1$
(sometimes $\Delta l = 3$ transitions are observed), and one generally
observes all the transitions in each supermultiplet. For example, the
multiplets $n\,^7\!D\odd - s^6D5d\,^M\!L$ (2347--2354, 2356 and 2357 )
are due to $5p-5d$ transitions between terms having the parent term
$3d^64s\ ^6\!D$, and if most of the lines in multiplet
$n\,^7\!D\odd-s^6D5d\ ^5\!D$ (2347) are seen then lines of the same or
greater intensity in multiplets $n\,^7\!D\odd-s^6D5d\ ^5\!G$ (2348),
$n\,^7\!D\odd-s^6D5d\ ^7\!F$ (2349), $n\,^7\!D\odd-s^6D5d\ ^7\!D$
(2350) and so on, should also be observed.

Highly-excited levels due to $3d^64s(^6D)nl$ odd-parity
subconfigurations combine with the $3d^7(^4F)4s$ subconfiguration as
well as the $3d^64s^2$ configuration, and the $3d-nl$ transitions are
seen in the same spectral region as the $4s-nl$ transitions because of
the similar energy of the $3d$ and $4s$ electrons in the transition
elements. These transitions can also be regarded as allowed
supermultiplets, because a possible alternative notation for
$3d^7(^4F)4s$ is $3d^64s(^6D)3d$. For example, multiplet
$a\,^5\!F-s^6D6p\ ^5\!F$ (165) is due to a $3d^64s(^6D)3d-3d^64s(^6D)6p$
transition. The lines in multiplets of $3d-nl$ transitions often have
irregular intensities due to the different coupling in the higher and
lower subconfigurations.

In some cases, configuration interaction gives multiplets between terms
with different parents - for example configuration interaction between
$3d^64s(^6D)5d$ and $3d^7(^4F)6s$ gives rise to the multiplet
$n\,^7\!D\odd - ^4\!F6s\ ^3\!F$ (2355), which is a transition between the
$3d^64s(^6D)5p$ and $3d^7(^4F)6s$ subconfigurations. In rather rare
cases, $\Delta l=3$ transitions occur between terms of the same parent:
the multiplet $a^3H-s\ ^6\!D_{4.5}\,4f$ (516) involves
transitions of the form $4s^2-4s4f$, which probably occur because of
configuration interaction between $3d^64s^2$ and $3d^74s$.

\section{Table of Multiplets}

The new multiplet table for Fe I is given in table 2. If a line is
present in the FT spectra, the wavenumber and intensity is taken from
these spectra; otherwise it is taken from the grating spectra. The criteria for
ordering the multiplets are the same as in the MT.
Each group of
multiplets consists of multiplets with the same lower term, and the
groups are ordered in increasing energy of the
lower term ($a\,^5\!D,\ a\,^5\!F,\ a\,^3\!F$ \dots). Within each group,
the multiplets are ordered in increasing energy of the upper term
($z\,^7\!D\odd, z\,^7\!F\odd, z\,^7\!P\odd$ \dots). It should be noted,
however, that some of the multiplets of Fe I listed in the MT
are not ordered according to these criteria, as Moore only
adopted the criteria part way through the typing of the Fe I table. This
means that the order of multiplets in Table 2 is not necessarily the
same as in the MT. Within each multiplet the lines have been
listed in wavelength order. This is partly so that the user can work
systematically through a spectrum and multiplet, and also because the
irregular intensities observed in many multiplets rendered less
meaningful the ordering scheme used in the MT -- where the main
diagonal, first satellites, and second satellites are listed in order of
decreasing J.

A more important difference between Table 2 and the MT is that we
present all the multiplets in one large table, which we consider to be
more convenient for the user. This means that it is impossible
to retain the old multiplet numbers of Moore, which in any case did not
extend to the infrared, where many of our new multiplets are to be
found. Our new numbers are given in column 1 of table 2. The term
designations are given in column 2, and underneath these designations are
given the number of the multiplet in the MT, where one exists, with the
usual prefix of `UV' for multiplet numbers taken from the UV multiplet
table (Moore 1950). We have indicated designations that have changed
since the MT with an asterisk after the number in the MT. Column 3 gives
the two J-values for the transition.

The decision as to what constitutes a multiplet is particularly awkward
where the levels are better described by the JK coupling scheme rather than
the LS coupling scheme. We have decided to group all lines due to the
same fine structure level of the parent, $J_c$, together. For example,
in multiplet $z\,^7\!D\odd - s^6D_{4.5}6d$ (454) all upper levels have the same
parent level, $3d^64s\,^6\!D_{9/2}$ in Fe II, but have different
K-values. For this reason we have labelled every line in these
multiplets with the abbreviated term designation, to avoid confusion
between levels of the same J-value but different $K$.

The intensity of the line is given in column 4. The intensities of lines
in the FT spectra are the logarithms of the integrated intensities and
are given to two decimal places. The intensities of lines observed only
in grating spectra are integers, which are on a different
scale to the FT spectra and are visual estimates of the photographic
blackening in the continuous hollow cathode spectra. Lines marked `d0'
are faint and diffuse, and those marked `d0?' are hardly detectable from
the background. Lines marked `b' are broad, and those marked with an
asterisk are blended with other unresolved lines. Where a line is not
present in the continuous hollow cathode spectrum, the intensity has
been taken from the pulsed hollow cathode and is given in italics in
the printed version of table 2 and in brackets in the computer-readable
version.

In the lower multiplets many lines are affected by self absorption or
self reversal. Self absorption widens the line and reduces its
intensity, but the wavenumber remains unaffected (Nave \etal 1991).
Intensities and experimentally measured wavenumbers are given for all of
these lines, but in almost all cases accurate gf values have been
measured and should be used instead of the intensities. Self-reversed
lines have a pronounced dip in the centre of the line profile, and
neither intensities nor accurate wavenumbers can be measured. No
intensities are given for self-reversed lines and the wavenumbers are
Ritz wavenumbers. All the FT intensities have been corrected
for the response of the spectrometers, but it was not possible to
maintain a completely consistent scale throughout the whole spectral
region covered by table 2 because spectra in different spectral regions
were obtained under several different source conditions, which affect
the relative intensities of the lines. Nevertheless the information is
still useful because it can help in identifying lines in other sources,
and in many cases it is the only alternative to a calculated gf value,
which can sometimes be wrong by an order of magnitude.

The vacuum wavelength given in column 5 is derived from the
measured wavenumber in column 8. Wavelengths and wavenumbers of
self-reversed and masked (i.e. obscured by a much stronger line) lines
are Ritz wavelengths and wavenumbers.
The superscript to the wavenumber is a measure of its uncertainty, taking into
account both the statistical and systematic errors described in section
2. The errors in wavenumbers of lines graded `A' are estimated to be
less than 0.005~\cm, lines graded `B' less than 0.01~\cm, lines graded
`C' less than 0.02~\cm, and lines graded `D' greater than 0.02~\cm. All
known blended lines and all lines measured only in grating spectra have
been assigned the grade `D'. The wavelength uncertainty $\Delta\lambda$
in \AA\ is obtained from the wavenumber uncertainty $\Delta\sigma$ in
\cm\ by using the relation:
\begin{equation}
\Delta\lambda = \frac{\Delta\sigma}{\sigma^2}\cdot10^{8}
              = \Delta\sigma\cdot\lambda^2\cdot10^{-8}
\end{equation}
and typical values are given in table 3.

\begin{planotable}{llllllll}
\tablewidth{0pt}
\tablenum{3}
\tablecaption{Uncertainties in wavenumbers and wavelengths}
\tablehead{
Grade &    & $\Delta\sigma$ & \multicolumn{5}{c}{ $\Delta\lambda ($m\AA)}\nl
      &    &   (\cm)        &2000~\AA\ &5000~\AA\ & 1~\mic & 2~\mic&5~\mic}
\startdata
A     &$<$ & 0.005          & 0.2      & 1.25     & 5      & 20    & 125 \nl
B     &$<$ & 0.01           & 0.4      & 2.5      & 10     & 40    & 250 \nl
C     &$<$ & 0.02           & 0.8      & 5.0      & 20     & 80    & 500 \nl
D     &$>$ & 0.02           & 0.8      & 5.0      & 20     & 80    & 500 \nl
\end{planotable}

The air wavelengths in column 6 for all lines above 2000~\AA\ have been
derived from the wavenumbers using the Edl\'en dispersion formula
(Edl\'en 1966):
\begin{eqnarray}\label{edlen}
\lefteqn{\lambda_R(air) = \nonumber} \\
 &&  \frac{10^8}{\sigma_R}\times  \left( 1 + 8342.13\times10^{-8}+
\frac{15997}{3.89\times10^9-\sigma_R^2} \right. \nonumber \\
 &&  \left. \hspace{4em} +
\frac{2406030}{1.3\times10^{10}-\sigma_R^2}\right)^{-1} \nonumber \\
\end{eqnarray}
which agrees to within 1~m\AA\ with more recent infrared
measurements by Peck and Reeder (1972). Column 7 gives the difference
between the observed wavelength and the Ritz wavelength in m\AA. To
obtain the Ritz wavelengths, the differences in column 7 should be
subtracted from the wavelengths in column 5 or 6. Column 9 gives the
difference between the observed wavenumber and the Ritz wavenumber, and
the Ritz wavenumbers should be determined       in a similar way. A large
discrepancy between the measured wavelength of a line graded `A' and the
Ritz wavelength is indicative of an unknown and unresolved blend. Lines
which are known to be blended with other Fe I, Fe II, Ne or Ar lines are
marked in column 13 with the species of the blended line. Lines which
are obscured by a much stronger line are marked with an `M' in column
13, together with the species of the stronger, masking line.

The wavenumbers of all lines are necessarily those emitted by the iron
atom in a specific set of plasma conditions, and are not those of an
isolated atom. A detailed discussion of possible wavenumber shifts is
given in Learner \& Thorne (1988). The majority of our lines with upper
levels of low excitation ($<$6~eV) have a very small Lorentzian
component to the line profile, and can thus be expected to be relatively
free from wavelength shifts. Lines graded `A' emitted from upper levels
of low excitation should be selected where the accuracy of the
wavelength is of particular importance, and the best of these are listed
in the three recently published tables of recommended wavelength
standards (Learner \& Thorne 1988, Nave \etal 1991, Nave \etal 1992).
Lines originating from levels of high excitation are Lorentz broadened
and may be subject to pressure shifts. A recent comparison of the $4f -
5g$ transitions observed in our spectra with those observed in the sun
indicated that the solar wavenumbers were roughly 0.006~\cm\ less than
the laboratory wavenumbers, which is probably due to pressure or
current-dependent shifts in the hollow cathode (Johansson \etal 1994).
Another possible source of wavenumber shifts is unresolved or
partly-resolved isotope structure. We have as yet seen no evidence for
isotope structure in Fe I, although it is of importance in Fe II and Ni
I (Rosberg, Johansson \& Litz\'en 1992, Litz\'en, Brault \& Thorne 1993).

Columns 10 and 11 contain the excitation potentials of the lower and
upper levels of the transition respectively. These are given in eV in
the printed version of table 2, and in eV and \cm\ in the
computer-readable version. Experimental gf values have been taken from various
sources in the literature and are given in column 12 (Fuhr, Martin \& Wiese
1990;
O'Brian \etal 1991; Meylan \etal 1993, Johansson \etal 1994). We have
not estimated the accuracy of these measurements, and the user is
advised to consult the original references. In particular, we note that
some publications lead to a solar abundance of iron which
is slightly greater than the currently accepted value (Holweger \etal
1991). Other recent measurements of gf-values or lifetimes in Fe I
include Kock \etal (1984), Blackwell \etal (1986), Bard \etal (1991),
and Engelke, Bard \& Kock (1993). We have not included calculated gf values for
the lines, partly because of their uncertain accuracy, but also because they
are likely to be re-calculated with the new levels included.
Calculated gf values for Fe I can be found in the tables of
semi-empirical gf values by Kurucz (1989) or Fawcett (1987). Although
the absolute accuracy of these calculations is often rather poor, the
relative accuracy of calculated gf values of lines within a single
multiplet can be quite good (Fawcett 1987, Blackwell 1983). The
calculations of the Opacity Project (Sawey \& Berrington 1992) include
multiplet oscillator strengths, which may be helpful in estimating gf
values for lines due to highly-excited levels that have not been
measured or calculated by other groups.

Table 4 is a finding list for all the lines in table 2. For all lines
above 2000~\AA\ we give an air wavelength, $\lambda_{air}$, a
wavenumber, $\sigma$, and the number of the multiplet in table 2 where
the line can be found. We have only given vacuum wavelengths
$\lambda_{vac}$ below 2000~\AA, as in the air region they can easily be
obtained from the wavenumber of the line:
\begin{equation}\label{stol}
\lambda_{vac}({\rm \AA})=10^8/\sigma({\rm cm}^{-1})
\end{equation}.

Table 5 gives 125 of the strongest unidentified lines in our spectra
with a signal-to-noise ratio greater than 100. These are all probably
due to Fe~I, as they are present in either the solar spectrum or the
absorption spectra of Brown \etal (1988). Some of these lines may well
be due to the unclassified levels in Brown \etal (1988), but we have
been unable to confirm them. The rest are probably due to highly-excited
levels that we have been unable to find.

\section{Fe I lines not listed in the table}

The MT contains lines and multiplets that are not listed in table 2.
Many of these lines have not been observed in any laboratory spectra, but
are lines predicted from energy levels for which the Ritz wavelength
coincides with a feature in the solar spectrum. They are designated in
the MT by the symbol \sun\ in the intensity column and the letter `P' in
the reference column. As we do not see them in our spectra, we are
unable to say whether all these lines are in fact due to Fe I. The Ritz
wavelengths in the MT were based upon old energy level values, and more
accurate wavelengths for them can be determined from the energy levels
listed in table 1, provided that the identifications are correct and are
unchanged. The Ritz wavenumber, $\sigma_R$ in \cm, is given by the
difference between the upper and lower level values of the transition,
and the Ritz vacuum wavelength in \AA\ can be calculated from equation
\ref{stol}. To obtain the Ritz air wavelength $\lambda_R(air)$, the
Edl\'en dispersion formula in equation \ref{edlen} should be used
(Edl\'en 1966).

Ritz wavenumbers and wavelengths for all Fe I lines are also listed in
the calculations of semi-empirical gf values by Kurucz (1989), and use
of Kurucz's list has the advantage that calculated gf values are also
given, which would assist in judging whether a particular solar or
stellar line is due to Fe I or not. It should be emphasised, however,
that the wavelengths listed in Kurucz's calculations will only be
correct for energy levels that have been determined experimentally. This
means that reliable wavelengths of lines will only be found in the
smaller linelists distributed by Kurucz. The larger linelist also
contains lines due to energy levels that have been calculated by atomic
structure computer programs, and the wavelengths of these lines may be
wrong by several \AA\ or more, depending on the wavelength and accuracy
of the calculations. The larger linelist should only be used when the
details of the spectrum are unimportant (e.g. for opacity and radiation
transport calculations) and {\em it cannot be used for high-resolution
spectroscopy}.

Two other publications also contain lists of Fe I lines that are not
present in table 2. Brown \etal (1988) have analysed the absorption
spectrum of Fe I in the range 1550~\AA\ to 3215~\AA. Roughly 800 of the
3000 lines they observed are below 1700~\AA, and fall outside the region
of table 2. Between 1700~\AA\ and 3215~\AA\ they also observed many
lines which are not present in our spectra. These are due to
highly-excited odd-parity levels which are not well populated in a
hollow cathode, but combine strongly with the ground term and
consequently give strong lines in an absorption spectrum. Brown \etal (1988)
list roughly 100 of these highly-excited levels. In the infra-red, Schoenfeld,
Chang \& Geller (1993) have identified two supermultiplets in the region of
3900~\cm\  (2.56~\mic) and 1350~\cm\ (7.41~\mic) due to $3d^64s(^6D)\
4f-6g$ and $3d^64s(^6D)\ 5g-6h$ respectively, which have been observed
in the solar spectrum. They also identify several features at longer
wavelengths which are due to higher Rydberg transitions. All but the
$4f-6g$ supermultiplet are outside the range of our emission spectra, and this
supermultiplet is not present due to the high excitation energy of the $6g$
levels.

\section{Summary}

The total number of energy levels in table 1 is 846, of which 28 are
new. In comparison, the total number of energy levels from which the
lines in the MT are derived is 467. Many of the new energy levels are of
high excitation and are likely to be particularly useful in the
interpretation of astrophysical spectra. The total number of lines in
table 2 is 9501, which are due to 9759 transitons arranged into 2785
multiplets. This compares with the $\sim$5500 lines in the MT, of which
roughly 1650 are Ritz wavelengths of lines present in the solar
spectrum, but not observed in laboratory spectra. The biggest increase
is in the ultraviolet below 3000\AA, where the number of lines has
increased from $\sim$750 to $\sim$2000, and in the infrared above 1\mic,
where none of the $\sim$3000 lines in table 2 were given in the MT. The
strongest lines have an uncertainty of $<$0.002\cm\ (0.2~m\AA\ at
3000~\AA\ and 8~m\AA\ at 2~\mic), which is up to an order of magnitude
better than wavelength standards derived from previous data (Nave \etal
1991). Almost all of the lines are present in the solar spectrum, and in
many cases correspond to strong lines.

Table 5 gives a good indication of the current state of the analysis of Fe
I. Almost all of the strongest lines in the visible have been
identified. In spite of the many new identifications in the infrared,
the largest proportion of unidentified lines fall between 1 and 2~\mic.
In the near UV, both our FT and grating spectra contain many
unidentified lines, many of which are also strong unidentified lines in
the solar spectrum. At present, no laboratory spectra of sufficient
quality beyond 5.4~\mic\ have been recorded, and it is not possible for
us to estimate the contribution of Fe I to stellar spectra in this
region. Table 2 also shows that relatively few lines in the infrared
beyond 1~\mic and few lines in the ultraviolet below 3000~\AA\ have
measured gf values. With the current interest in both ultraviolet and
infrared astronomy they will almost certainly be required for stellar
and solar spectroscopy.

Many users will prefer to have the new multiplet table in
computer-readable form. For the time being, ascii versions of tables 1 and 2
are available by anonymous FTP from {\tt ferrum.fysik.lu.se}
(130.235.92.170) at Lund University, in the directory {\tt
pub/iron}. The ascii version of table 2 is also available sorted
in wavelength order.

\acknowledgements

G. Nave gratefully acknowledges a fellowship from the European Space
Agency. The work was supported by the U.K. Science and Engineering
Research Council, the Swedish National Space Board, and the Swedish
Natural Science Research Council. R. C. M. Learner wishes to thank the
NSO for provision of facilities and for assistance in recording the
visible and infrared FT spectra whilst he was there as a visiting
researcher. S. Johansson is grateful to the Atomic Spectroscopy group at
NIST for assistance in connection with the recordings of the grating
spectra. We are grateful to C.~J.~Harris for measurements of the UV FTS
spectra and to G.~Cox for the development of the hollow cathodes.

\end{document}